\documentclass[submission, Proceedings]{SciPost}

\hypersetup{
    colorlinks,
    linkcolor={red!50!black},
    citecolor={blue!50!black},
    urlcolor={blue!80!black}
}

\usepackage[bitstream-charter]{mathdesign}
\DeclareSymbolFont{usualmathcal}{OMS}{cmsy}{m}{n}
\DeclareSymbolFontAlphabet{\mathcal}{usualmathcal}

\begin{document}

% TODO: write your article's title here.
% The article title is centered, Large boldface, and should fit in two lines
\begin{center}{\Large \textbf{
BFKL phenomenology:
resummation of high-energy logs in inclusive processes\\
}}\end{center}

% TODO: write the author list here. Use initials + surname format.
% Separate subsequent authors by a comma, omit comma at the end of the list.
% Mark the corresponding author with a superscript *.
\begin{center}
	F.~G.~Celiberto~$^{1,2,3}$,
	M.~Fucilla~$^{4}$,
	D.~Yu.~Ivanov~$^{5}$,\\
	M.~M.~A.~Mohammed~$^{4,*}$,
	and A.~Papa~$^{4}$
\end{center}

% TODO: write all affiliations here.
% Format: institute, city, country
\begin{center}
	
	\centerline{${}^1$ {\sl European Centre for Theoretical Studies in Nuclear Physics and Related Areas (ECT*),}}
	\centerline{\sl I-38123 Villazzano, Trento, Italy}
	\vskip .18cm
	\centerline{${}^2$ {\sl Fondazione Bruno Kessler (FBK), %}}
			%\centerline{\sl
			I-38123 Povo, Trento, Italy} }
	\vskip .18cm
	\centerline{${}^3$ {\sl INFN-TIFPA Trento Institute of Fundamental Physics and Applications,}}
	\centerline{\sl I-38123 Povo, Trento, Italy}
	\vskip .18cm
	\centerline{${}^4$ {\sl Dipartimento di Fisica, Universit\`a della Calabria,}}
	\centerline{{\sl and Istituto Nazionale di Fisica Nucleare, Gruppo collegato
			di Cosenza,}}
	\centerline{\sl I-87036 Arcavacata di Rende, Cosenza, Italy}
	\vskip .18cm
	\centerline{${}^5$ {\sl Sobolev Institute of Mathematics, 630090 Novosibirsk,
			Russia}}
	\vskip .18cm
	% TODO: provide email address of corresponding author
	* mohammed.maher@unical.it
\end{center}

\begin{center}
\today
\end{center}

% For convenience during refereeing (optional),
% you can turn on line numbers by uncommenting the next line:
%\linenumbers
% You should run LaTeX twice in order for the line numbers to appear.

\definecolor{palegray}{gray}{0.95}
\begin{center}
\colorbox{palegray}{
  \begin{tabular}{rr}
  \begin{minipage}{0.1\textwidth}
    \includegraphics[width=30mm]{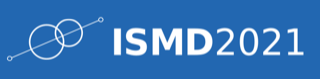}
  \end{minipage}
  &
  \begin{minipage}{0.75\textwidth}
    \begin{center}
    {\it 50th International Symposium on Multiparticle Dynamics}\\ {\it (ISMD2021)}\\
    {\it 12-16 July 2021} \\
    \doi{10.21468/SciPostPhysProc.?}\\
    \end{center}
  \end{minipage}
\end{tabular}
}
\end{center}

\section*{Abstract}
{\bf
% TODO: write your abstract here.
We present recent phenomenological studies, tailored on kinematic configurations typical of current and forthcoming analyses at the LHC, for two novel probe channels of the BFKL resummation of energy logarithms. Particular attention is drawn to the behavior of distributions differential in azimuthal angle and rapidity, where significant high-energy effects are expected.
}

% TODO: include a table of contents (optional)
% Guideline: if your paper is longer that 6 pages, include a TOC
% To remove the TOC, simply cut the following block
\vspace{10pt}
\noindent\rule{\textwidth}{1pt}
\tableofcontents\thispagestyle{fancy}
\noindent\rule{\textwidth}{1pt}
\vspace{10pt}

\section{Introduction}
\label{sec:intro}
% TODO: write your article here.
With more and more data to be collected at the Large Hadron Collider (LHC), the study of semi-hard processes~\cite{Gribov:1983ivg} in the large center-of-mass energy limit gives us an opportunity to further test perturbative QCD (pQCD) in unexplored kinematical configurations thus contributing to a better understanding of the dynamics of strong interactions. Within pQCD computations, reducing theoretical uncertainties coming from higher order corrections is required to have a reliable estimate of the production rate. At high energies, the validity of the perturbative expansion, truncated at a certain order in the strong coupling $\alpha_s$, is spoiled. This is due to the appearance of large logarithms of the center-of-mass energy squared, $s$, associated with the perturbative calculations  and it is needed to resum them to all orders in $\alpha_{s}$. The most powerful framework to perform this resummation is the Balitsky–Fadin–Kuraev–Lipatov (BFKL)~\cite{Fadin:1975cb,kuraev1976multi,Kuraev:1977fs,Balitsky:1978ic} approach, initially developed at the so called leading logarithmic approximation (LLA), where it prescribes how to resum all terms proportional to $(\alpha_s \ln s )^n$. In order to improve the obtained results at LLA, the so called next-to-leading logarithmic approximation (NLA) was considered~\cite{Fadin:1998py,Ciafaloni:1998gs}, where also all terms proportional to $\alpha_s(\alpha_s \ln s)^n$, were resumed. Clearly, a significant question for collider phenomenology is highlighting at which energies the BFKL dynamics becomes significant and cannot be overlooked. Typical BFKL observables that can be studied at the LHC are the azimuthal coefficients of the Fourier expansion of the cross section differential in the variables of the tagged objects over the relative azimuthal-angle. They take a certain factorization form given as the convolution of a universal BFKL Green’s function with process dependent impact factors, the latter describing the transition from each colliding proton to the respective final-state identified object. The BFKL Green's function obeys an integral equation, whose kernel is known at the next-to-leading order (NLO)~\cite{Fadin:1998py,Ciafaloni:1998gs,Fadin:1998jv,Fadin:2004zq,Fadin:2005zj}.  
Over last years, pursuing the goal of identifying observables that fit the data where BFKL approach is needed, a number of reactions have been proposed for different collider environments: the exclusive diffractive leptoproduction of two light vector mesons~\cite{Pire:2005ic,Segond:2007fj,Enberg:2005eq,Ivanov:2005gn,Ivanov:2006gt}, the inclusive hadroproduction of two jets featuring large transverse momenta and well separated in rapidity, the so-called Mueller-–Navelet jets~\cite{Mueller:1986ey}, for which several phenomenological studies have appeared during last years~\cite{Colferai:2010wu,Caporale:2012ih,Ducloue:2013hia,Ducloue:2013bva,Caporale:2013uva,Caporale:2014gpa,Colferai:2015zfa,Caporale:2015uva,Ducloue:2015jba,Celiberto:2015yba,Celiberto:2015mpa,Celiberto:2016ygs,Celiberto:2016vva,Caporale:2018qnm}, the inclusive detection of two light-charged rapidity-separated hadrons~\cite{Celiberto:2016hae,Celiberto:2016zgb,Celiberto:2017ptm}, three- and four-jet hadroproduction~\cite{Caporale:2016zkc,Caporale:2016xku}, $J/\Psi$-plus-jet~\cite{Boussarie:2017oae}, hadron-plus-jet~\cite{Bolognino:2019cac}, heavy-flavor~\cite{Bolognino:2021mrc,Bolognino:2021hxx,Celiberto:2021dzy,Celiberto:2021fdp} and forward Drell–Yan dilepton production with a possible backward-jet tag~\cite{Golec-Biernat:2018kem}. The second class of probes for BFKL is given by single forward emissions in lepton-proton or proton-proton scatterings, giving us the possibility to probe the unintegrated
gluon distribution in the proton (UGD), which is linked to BFKL via the convolution between the BFKL gluon
Green’s function and the proton impact factor. Proposed channels to study the UGD are the exclusive light vector-meson electroproduction~\cite{Bolognino:2018rhb,Bolognino:2018mlw,Bolognino:2019bko,Bolognino:2019pba,Celiberto:2019slj,Bolognino:2021niq,Bolognino:2021gjm}, the exclusive quarkonium photoproduction~\cite{Bautista:2016xnp,Garcia:2019tne,Hentschinski:2020yfm}, and the inclusive tag of Drell–Yan pairs in forward directions~\cite{Motyka:2014lya,Brzeminski:2016lwh,Motyka:2016lta,Celiberto:2018muu}.

In this work we concentrate on the BFKL $\phi$-summed cross sections for two proposed reactions, the inclusive production of Higgs-plus-jet and $\Lambda$-plus-jet\footnote{The diffractive production of $\Lambda$-jet was studied by three of us~\cite{Celiberto:2020rxb}, with some predictions tailored on the CMS and CASTOR
typical kinematic ranges.} at the LHC, where the final tagged particles are well separated in rapidity. 
\section{Theoretical framework}
\label{sec:theory}
\begin{figure}[h]
\centering
\includegraphics[width=0.4\textwidth]{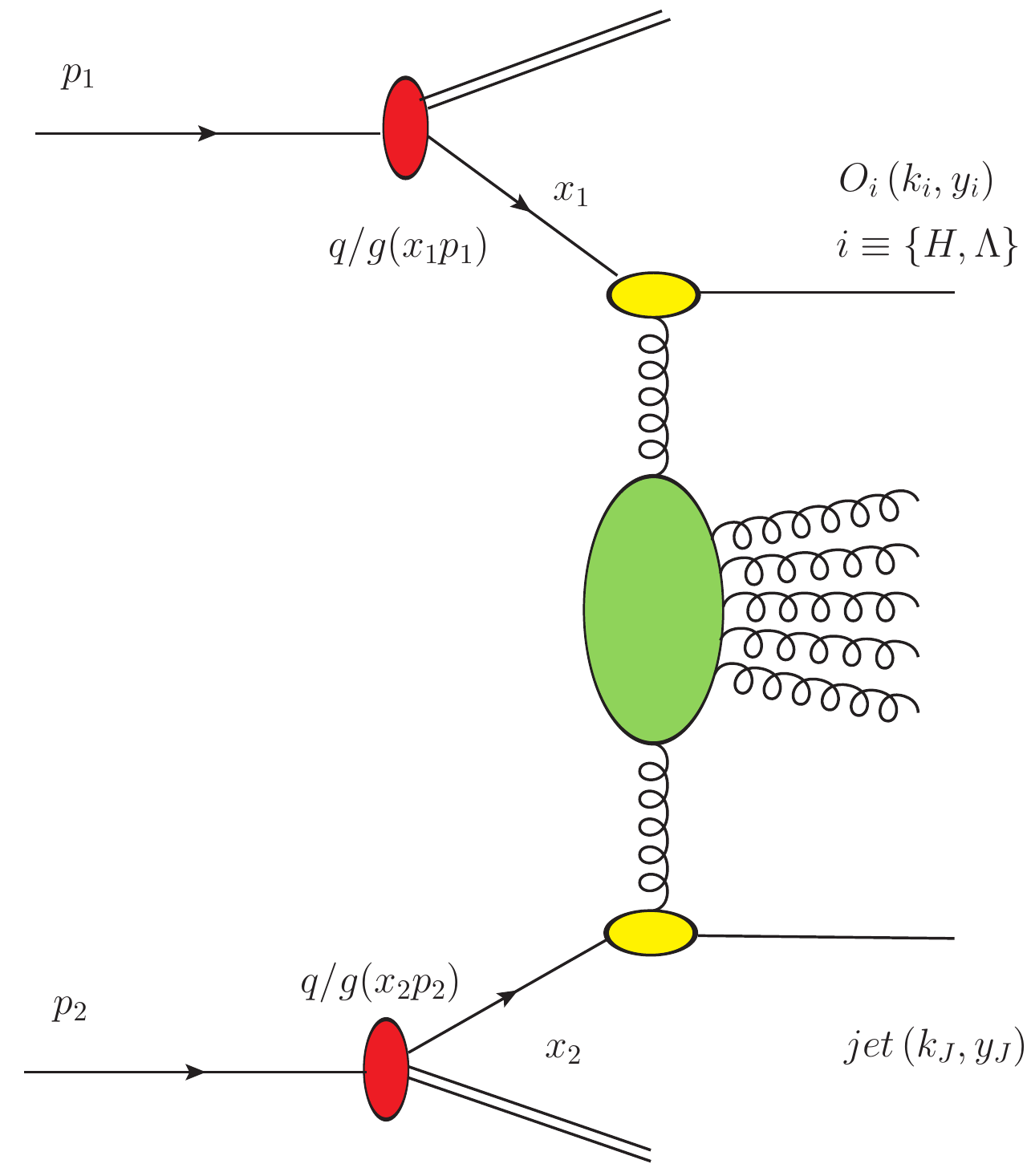}
\caption{Schematic representation of our considered semi-hard processes.}
\label{fig:semi-hard_processes}
\end{figure}

We present a general expression for the inclusive hadroproduction processes of our considerations~(depicted in Fig.~\ref{fig:semi-hard_processes}):
\begin{equation}\label{eq:semi_hard_process}
{\rm proton}(p_1) \ + \ {\rm proton}(p_2) \ \to \ {\rm O_i}(\vec k_i, y_i) \ + \ {\rm X} \ + \ {\rm jet}(\vec k_J, y_J),
\end{equation}
where a jet is always detected in association with the Higgs boson or the $\Lambda$-hyperon ($ {\rm O_i}(\vec k_i, y_i), i\\ \equiv \{{\rm H},\Lambda\}$), emitted with high transverse momenta $|\vec k_{i,J}|\equiv \kappa_{i,J}\gg \Lambda_{QCD}$, and large rapidity separation $\Delta Y = |y_i-y_J|$. The symbol $X$ stands for an undetected system of hadrons.
In the BFKL approach the cross section of the hard subprocesses can be presented as the Fourier sum of the azimuthal coefficients ${\cal C}_n$, 
having so
\begin{equation}\label{eq:BFKL_crssec}
\frac{d\sigma}
{dy_idy_J\, d\kappa_i \, d\kappa_Jd\phi_i d\phi_J}
=\frac{1}{(2\pi)^2}\left[{\cal C}_0+\sum_{n=1}^\infty  2\cos (n\phi )\,
{\cal C}_n\right]\, ,
\end{equation}
where $\phi=\phi_i-\phi_J-\pi$, with $\phi_{i,J}$ representing the Higgs/$\Lambda$ and jet azimuthal angles, while $y_{i,J}$ and $\kappa_{i,J}$ are their
rapidities and transverse momenta, respectively. The $C_{0}$ coefficient gives us the $\phi$-summed cross section, while the $C_{n\ne 0}$ ones are connected to the so-called azimuthal-correlation coefficients.

\section{Numerical analysis and discussion}
\label{sec:Numerical}
In order to match the realistic kinematic cuts adopted by the current experimental analyses
at the LHC, we integrate the coefficients $C_0$ over the phase space for the two
emitted objects, while their rapidity distance $\Delta Y$, is kept fixed
\begin{equation}\label{Integrated_coefficients}
C_0(\Delta Y,s) =
\int_{\kappa^{\rm min}_{i=H,\Lambda}}^{{\kappa^{\rm max}_{i=H,\Lambda}}}d|\vec \kappa_{i}|
\int_{\kappa^{\rm min}_J}^{{\kappa^{\rm max}_J}}d|\vec \kappa_J|
\int_{y^{\rm min}_{i=H,\Lambda}}^{y^{\rm max}_{i=H,\Lambda}}dy_{i}
\int_{y^{\rm min}_J}^{y^{\rm max}_J}dy_J
\delta \left( y_i - y_J - \Delta Y \right){\cal C}_0.
\end{equation}
We allow for a larger rapidity range of the jet in our two considered processes, $| y_J |< 4.7$, and  we consider asymmetric kinematic cuts for the final-state transverse momenta in ranges 10 GeV $< \kappa_{H} < 2m_{top}$ and 20 GeV $< \kappa_{J} < $ 60 GeV  for Higgs-jet, with $|y_H | < 2.5$ inside the CMS rapidity acceptances, while for the $\Lambda$ + jet case, the $\Lambda$-particle is considered to be detected in symmetric rapidity range: $-2.0$ to $2.0$ and transverse momenta (typical CMS measurements): 10 GeV  $<\kappa_\Lambda<$ 21.5 GeV and 35 GeV $< \kappa_J <$ 60 GeV. All numerical simulations were done using the hybrid Fortran2008/Python3 modular package {\tt JETHAD}~\cite{Celiberto:2020wpk}. The MMHT~2014~PDF set were used via the Les Houches Accord PDF Interface (LHAPDF) 6.2.1~\cite{Buckley:2014ana}, together with the AKK~2008~\cite{Albino:2008fy} FF interfaces which describe the hadronization probability of the initial state partons into the final state detected $\Lambda$ hyperon.

\begin{figure}[h]
	\centering
	%\hspace*{-0.8cm}
	\includegraphics[width=0.45\textwidth]{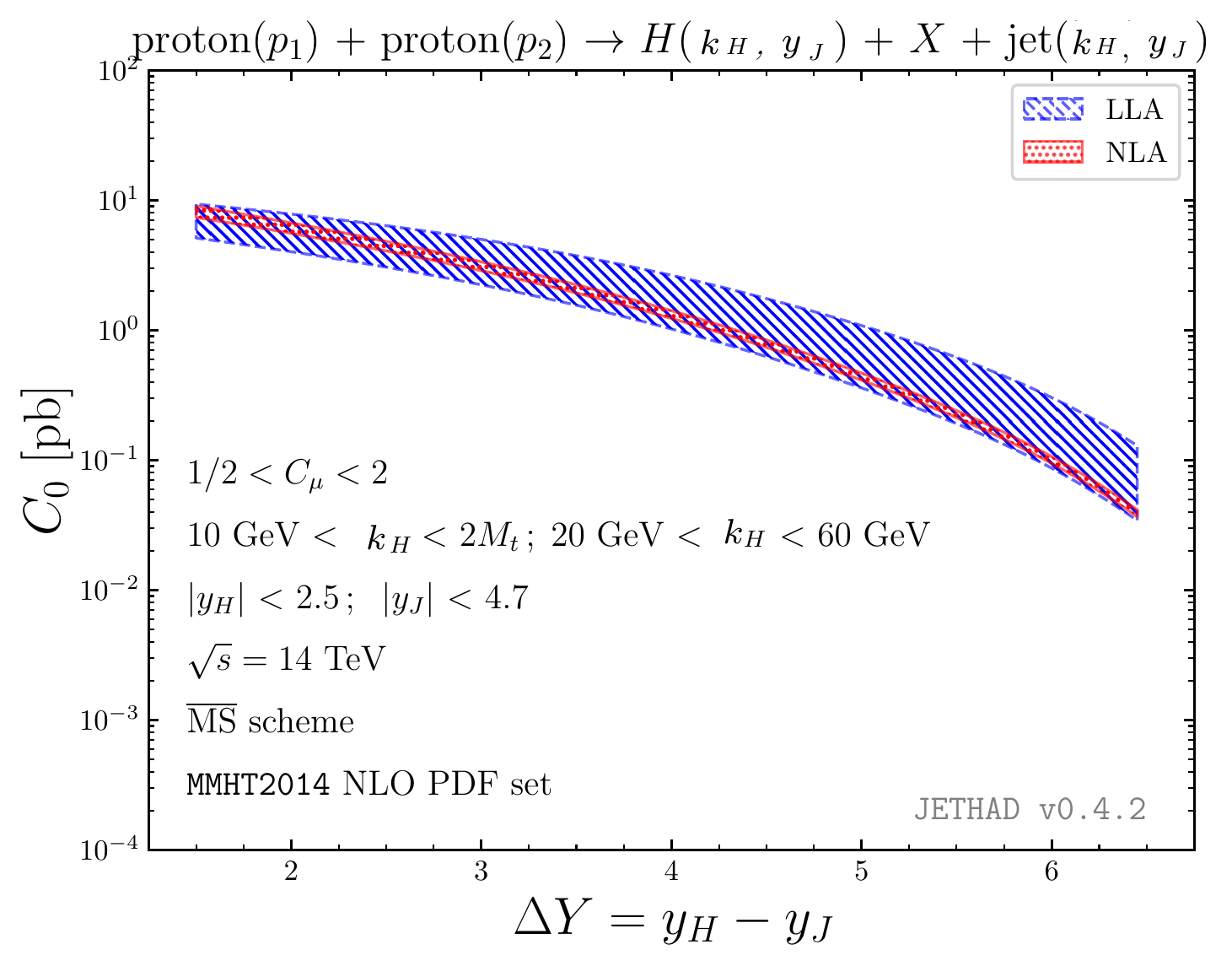}
	\includegraphics[width=0.45\textwidth]{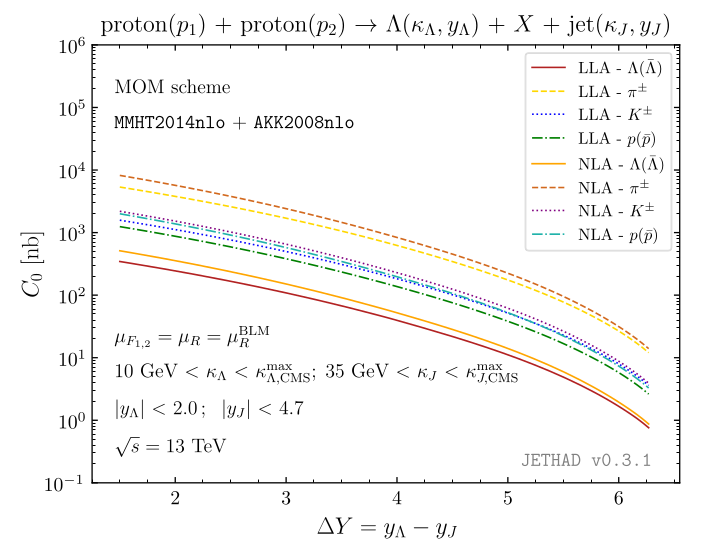}  
	\caption{$\Delta Y$-dependence of the $C_0$ in the symmetric $p_T$-range, for the inclusive Higgs-jet (left plot), and $\Lambda$-jet process (right plot), with  $\kappa^{max}_{\Lambda,CMS}=$ 21.5 GeV, and $\kappa^{max}_{J,CMS} =$ 60 GeV. }
	\label{fig:C0_0J}
\end{figure}

In Fig.~\ref{fig:C0_0J} we present results for the $\Delta Y$-dependence of the $\phi$-summed cross section, in the asymmetric kinematic configurations for both Higgs + jet (left plot) and the $\Lambda$ + jet (right plot) reactions. Here, the usual behavior of BFKL effects comes easily into play, where the growth with energy of the pure hard cross section as an effect of the resummation of high-energy logarithms, is downtrend by the convoluted PDFs and FFs. For the Higgs-jet process, we can remarkably notice that NLA predictions (red) are almost entirely nested inside LLA uncertainty bands (blue), since the large energy scales provided by the emission of a Higgs boson suppress the higher order corrections~\cite{Celiberto:2020tmb,Celiberto:2021tky}.
Cross sections in the $\Lambda$-jet channel are steadily lower when compered to 
the previously studied di-hadron~\cite{Celiberto:2016hae,Celiberto:2016zgb,Celiberto:2017ptm} and the hadron-jet~\cite{Bolognino:2019cac} reactions, this, together with the fact that the lower cutoff to identify $\Lambda$ hyperon is 10 GeV, which is larger than the
corresponding one for any light-hadron tagging, that gives us the opportunity to quench the experimental minimum-bias effects.

\section{Conclusion}
We have proposed two inclusive hadroproduction reactions, the Higgs boson plus a jet, and $\Lambda$-particle in association with a jet, as another novel
diffractive semi-hard channels to test the BFKL resummation. In both cases the final detected particles feature high transverse momenta and separated by a large rapidity distance. At variance with previously studied reactions, the Higgs + jet production channel exhibits quite a fair stability under higher-order corrections, and $\Lambda$-particle emissions in the final state dampen the experimental minimum-bias contamination, thus easing the comparison with forthcoming LHC data.

The next step in our program of investigating semi-hard phenomenology consists in performing the full NLA BFKL analysis for the Higgs + jet channel, via including the full NLO jet and Higgs impact factors, and extend our study to cover the kinematic configurations for the new-generations of colliders, such as HL-LHC~\cite{Chapon:2020heu}, EIC~\cite{AbdulKhalek:2021gbh}, NICA~\cite{Arbuzov:2020cqg}, and the FPF~\cite{Anchordoqui:2021ghd}.

\bibliography{References}

%\bibliography{SciPost_Example_BiBTeX_File.bib}

\nolinenumbers

\end{document}